\documentclass[twocolumn]{aastex631}

\shorttitle{The build-up of low mass passive galaxies  since  $z\sim3$}
\shortauthors{Santini et al.}
\graphicspath{{./}{figs/}}

\begin{document}

\title{The Stellar Mass Function in CANDELS and Frontier Fields: the build-up of low mass passive galaxies  since $z\sim3$ \footnote{Released on ...}}

\author[0000-0002-9334-8705]{Paola Santini}
\affiliation{INAF - Osservatorio Astronomico di Roma, \\
via di Frascati 33, \\
00078 Monte Porzio Catone, Italy}

\author{Marco Castellano}
\affiliation{INAF - Osservatorio Astronomico di Roma, \\
via di Frascati 33, \\
00078 Monte Porzio Catone, Italy}

\author{Adriano Fontana}
\affiliation{INAF - Osservatorio Astronomico di Roma, \\
via di Frascati 33, \\
00078 Monte Porzio Catone, Italy}

\author{Flaminia Fortuni}
\affiliation{INAF - Osservatorio Astronomico di Roma, \\
via di Frascati 33, \\
00078 Monte Porzio Catone, Italy}

\author{Nicola Menci}
\affiliation{INAF - Osservatorio Astronomico di Roma, \\
via di Frascati 33, \\
00078 Monte Porzio Catone, Italy}

\author{Emiliano Merlin}
\affiliation{INAF - Osservatorio Astronomico di Roma, \\
via di Frascati 33, \\
00078 Monte Porzio Catone, Italy}

\author{Amanda Pagul}
\affiliation{Department of Physics and Astronomy, University of California Riverside, \\
Pierce Hall, Riverside, CA., USA}

\author{Vincenzo Testa}
\affiliation{INAF - Osservatorio Astronomico di Roma, \\
via di Frascati 33, \\
00078 Monte Porzio Catone, Italy}

\author{Antonello Calabr\`o}
\affiliation{INAF - Osservatorio Astronomico di Roma, \\
via di Frascati 33, \\
00078 Monte Porzio Catone, Italy}

\author{Diego Paris}
\affiliation{INAF - Osservatorio Astronomico di Roma, \\
via di Frascati 33, \\
00078 Monte Porzio Catone, Italy}

\author{Laura Pentericci}
\affiliation{INAF - Osservatorio Astronomico di Roma, \\
via di Frascati 33, \\
00078 Monte Porzio Catone, Italy}



\begin{abstract}

  Despite significant efforts in the recent years, the physical
  processes responsible for the formation of passive galaxies through
  cosmic time remain unclear. The shape and evolution of the Stellar
  Mass Function (SMF) give an insight into these mechanisms. Taking
  advantage from the CANDELS and the deep Hubble Frontier Fields (HFF)
  programs, we estimated the SMF of total, star-forming and passive
  galaxies from $z=0.25$ to $z=2.75$ to unprecedented depth, and focus
  on the latter  population. The density of passive galaxies underwent a
  significant evolution over the last 11 Gyr.  They account for 60\%
  of the total mass in the nearby Universe against $\sim$20\% observed
  at $z\sim 2.5$.  The inclusion of the HFF program allows us to
  detect, for the first time at $z>1.5$, the characteristic upturn in
  the  SMF of passive galaxies at low masses, usually associated with environmental
  quenching.  We observe two separate populations of passive galaxies
  evolving on different timescales: roughly half of the high mass
  systems were already quenched at high redshift, while low mass
  passive galaxies are gradually building-up over the redshift range
  probed.  In the framework of environmental-quenching at low masses, we interpret this finding as evidence of 
  an increasing role
  of the environment in  the build-up of passive galaxies   as a function of
  time. Finally, we compared our findings with a set of theoretical
  predictions. Despite good agreement in some redshift and mass
  intervals, none of the models are able to fully reproduce the
  observations. This calls for further investigation into the involved
  physical mechanisms, both theoretically and observationally,
  especially with the  brand new  JWST data.

\end{abstract}



\section{Introduction}\label{sec:intro}

Galaxies are divided into two broad populations: star-forming
galaxies, hosting on-going star formation processes, and passive or
quiescent galaxies, characterized by negligible levels of star
formation rate (SFR) and evolving only through the aging of their
stellar populations.  \cite{peng10} identified two routes for
quenching, i.e. two classes of processes responsible for suppressing
the star formation and transforming star-forming galaxies into
passive.  The first quenching mode, also referred to as
``mass-quenching'', is causally linked to the star formation activity,
hence directly depending on the galaxy mass through the Main Sequence
relation \citep[e.g.][]{brinchmann04,elbaz07,santini09,santini17}. It
is mostly explained by processes involving gas heating through
supernovae or gas removal through AGN feedback.  The second mode,
dubbed ``environmental-quenching'', is attributed to processes related
to the effect of the galaxy environment, 
satellite quenching
such as ram pressure stripping \citep{gunn72},  galaxy harassment \citep{farouki81,moore96}, strangulation \citep{larson80,peng15}, 
 and mergers 
\citep{toomre72,springel05c}.  Environmental-related processes and subsequent
galaxy quenching have indeed been observed to be in place in clusters
\citep[e.g.][]{poggianti17,brown17}.  A detailed discussion on the
different modes of galaxy quenching can be found in \cite{man18}.

These two quenching modes show a differential behavior with the galaxy
stellar mass, with mass-quenching dominating at high masses and
environmental-quenching becoming increasingly important at low masses
\citep{peng10,peng12,geha12,quadri12,xie20}. It has therefore been suggested
that the two effects leave an imprint in the SMF, that is usually fit
well by two separate Schechter functions
\citep{schechter76}. According to the model of \cite{peng10}, while
mass-quenching is responsible for the dominant  Schechter component, showing a
similar characteristic mass as the SMF of star-forming galaxies  but with a
shallower slope, environmental-quenching produces a secondary
component with a steeper slope. This secondary Schechter component has
been clearly identified in the local Universe
\citep[e.g.][]{pozzetti10,baldry12}, and then at progressively higher
redshift up to $z\lesssim 1.5$ as the surveys became deeper
\citep[e.g.][]{tomczak14,mortlock15,davidzon17,mcleod21}. However,  these current
estimates of the  SMF  of the passive galaxy population do not push to stellar masses low enough
to identify it at earlier epochs.

In this work we investigate the evolution of the secondary component
of the SMF of passive galaxies, leading to an upturn at low masses, up
to $z\lesssim3$.  We take 
advantage of  the combination of the CANDELS survey and the deep HST Frontier Fields
observations,  and adopt 
a robust technique to estimate the SMF and
correct for incompleteness effects. 
 The paper is organized as follows. We illustrate the data set and methodology in Sect.~\ref{sec:data}; we present our results in  Sect.~\ref{sec:results}; in Sect.~\ref{sec:disc} we discuss their physical interpretation and compare them with theoretical predictions; we conclude in Sect.~\ref{sec:summary}. 
In the following, we adopt the
$\Lambda$ Cold Dark Matter ($\Lambda$CDM) concordance cosmological
model ($H_0 = 70$ km s$^{-1}$ Mpc$^{-1}$, $\Omega_M = 0.3$, and
$\Omega_\Lambda = 0.7$) and a \cite{chabrier03} Initial Mass Function
(IMF). All magnitudes are in the AB system.

\section{Data set and methods}\label{sec:data}

\begin{figure*}[t!]
    \centering
\includegraphics[width=0.95\columnwidth,clip,viewport=120 25 555
750,angle=270]{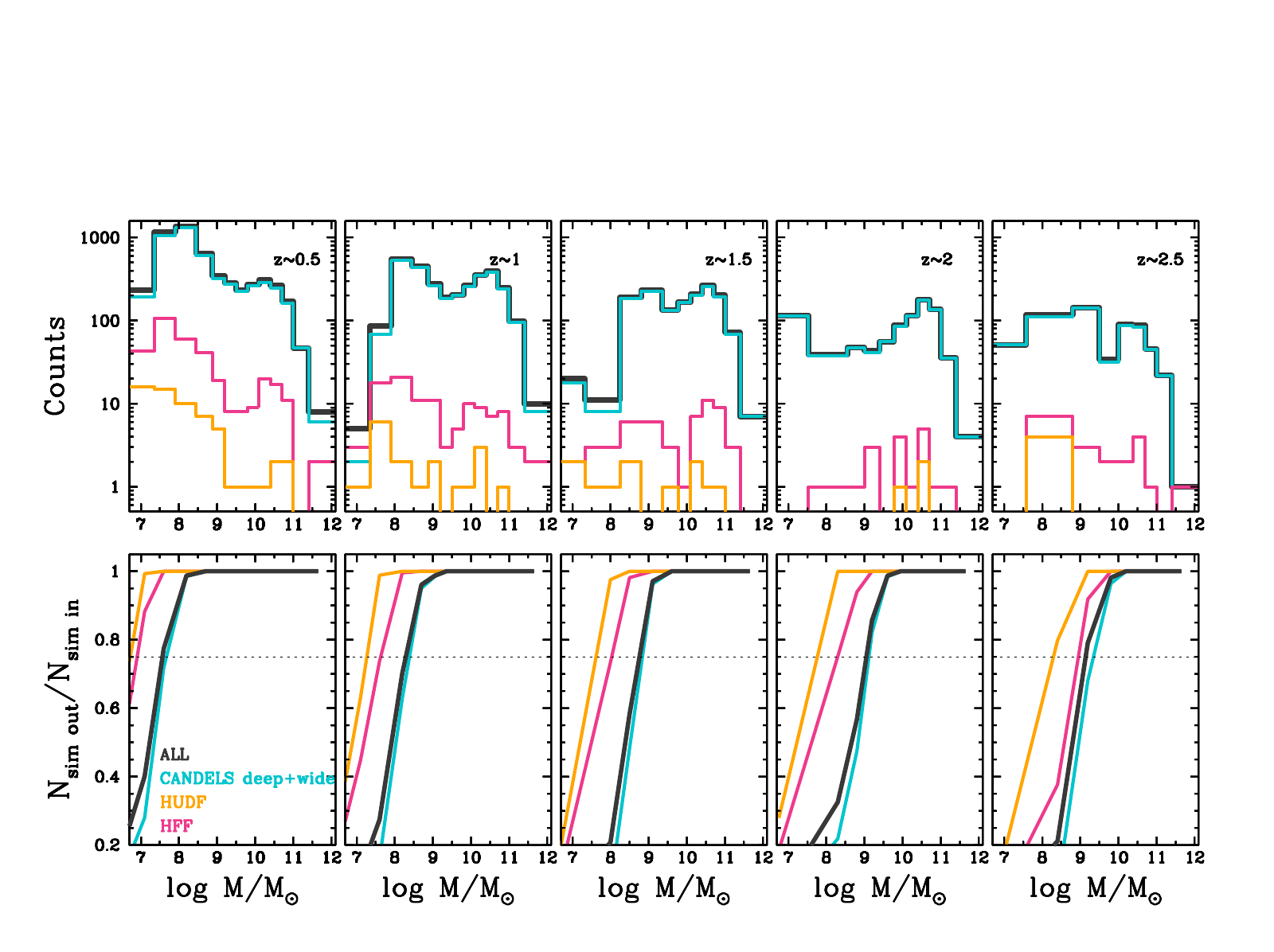}
\caption{{\it Upper} panels: Mass counts of passive galaxies in different redshift bins in the CANDELS deep and wide sample (i.e. all CANDELS except the HUDF area, turquoise), in HUDF (orange), in the HFF data set (magenta), and  in the total sample (black). {\it Lower} panels: Observational completeness of passive galaxies calculated by means of  simulations for the different data sets (same color coding as above). The horizontal dotted lines mark the 75\% completeness above which the SMF is computed.}
              \label{fig:counts}
\end{figure*}

\subsection{Photometric catalogs}

Our sample comprises the CANDELS \citep{koekemoer11,grogin11} and the
Hubble Frontier Fields \citep[HFF,][]{lotz17} programs. Even though
CANDELS, with its almost 1000 arcmin$^2$, represents a good compromise
in terms of area and depth, HFF observations allow us to push the
analysis of the SMF further down in stellar mass  thanks to the increased statistics at faint magnitudes.  

We used the
official photometric catalogs
\citep{barro19,galametz13,stefanon17,nayyeri17} and the latest
photometric redshift release of \cite{kodra22}  for all
CANDELS fields except GOODS-S, for which we adopted the new 43-bands
catalog and photo-$z$ of \cite{merlin21}. 

For the HFF we only
considered the parallel fields, and
used the multiwavelength catalogs, photometric redshifts and lensing
factors of \citealt{merlin16}, \citealt{castellano16} and
\citealt{dicriscienzo17}, assembled within the ASTRODEEP project, for
A2744, M0416, M0717 and M1149, and of \cite{pagul21} for
AS1063 and A370.  Given the modest lensing amplification in the parallel fields, we considered median magnification values for each field, ranging between 1 and 15\% (semi-interquartile ranges cover the interval 3-10\%).

 The catalogs used in this analysis comprise photometry in 43 (GOODS-S and COSMOS), 18 (GOODS-N), 19 (UDS), 23 (EGS) and 10 (HFF) bands. 
 To homogenize the sample as much as possible, for the AS1063 and A370 HFF catalogs, we used the same bands available for the other 4 fields. The $K$ band and IRAC CH1 and CH2 are available for all catalogs, while the CANDELS sample also includes CH3 and CH4.

\subsection{Stellar masses}

Stellar masses were estimated by means of SED fitting through the
proprietary code \textsc{zphot} \citep{fontana00}.

We assumed \cite{bc03} stellar population models, a \cite{chabrier03} IMF and delayed star formation histories (SFH($t$)$\propto (t^2/\tau) \cdot \exp(-t/\tau)$). The timescale $\tau$ of the declining exponential tail of the SFH ranges from 100 Myr to 7 Gyr, the age can vary between 10 Myr and the age of the Universe at each galaxy redshift, and metallicity can be 0.02, 0.2, 1 and 2.5 times Solar. We assumed a \cite{calzetti00} extinction law with E(B-V) varying from 0 to 1.1.
Nebular emission is included following the prescriptions of \cite{castellano14} and \cite{schaerer09}.

\subsection{Sample selections}
 
To ensure reliability of the inferred stellar masses, we cut the
sample at a signal to noise  ratio  (SNR) in the H160 band of at least 5.
We report in Table~\ref{tab:depths} the average total magnitudes corresponding to the adopted SNR threshold in the various subsets, calculated as the median magnitudes of sources with SNR between 4.8 and 5.2.

The sample was cleaned by removing sources affected by photometric issues,
X-ray selected AGNs as classified in the CANDELS official catalogs,
and stars.  These were identified either spectroscopically or
photometrically. In the latter case, we removed all sources with
SNR(H160)$>$10 and either {\it a)} SExtractor \citep{bertin96}
\textsc{class$\_$star}$>$0.95 or {\it b)} \textsc{class$\_$star}$>$0.8
and populating the stellar locus of the {\it BzK} diagram
\citep{daddi04}.

\subsection{The selection of passive and star-forming galaxies}

Passive and star-forming galaxies were identified from their rest-frame
($U$-$V$) and ($V$-$J$) colors ({\it UVJ} in the following, \citealt{williams09}). The {\it UVJ} selection is widely adopted in the literature and in previous SMF studies \citep[e.g.][]{muzzin13,tomczak14,mcleod21}. It was shown to agree with a selection based on the specific SFR \citep[e.g.][]{williams09,carnall18,carnall19}, and the analysis of \cite{whitaker13} demonstrated the reliability of {\it UVJ} selected quiescent galaxies by means of 3D-HST spectroscopy.

In this work, we adopted the redshift-dependent {\it
  UVJ} cuts of \cite{whitaker11}.   Since the {\it UVJ} technique is known to
be incomplete in terms of galaxies that have been recently quenched
\citep[][]{merlin18,schreiber18c,deshmukh18,carnall20}, whose abundance is
not negligible at $z\gtrsim 3$ due to the short timescales available
in the early Universe,  we limit our analysis 
analysis 
at $z<2.75$. 
  
Tables~\ref{tab:smf1} to \ref{tab:smf5} show the  number of galaxies in each redshift and mass bin for each galaxy population (total, passive and star-forming galaxies). The overall 
number of sources 
in each
redshift interval is listed in Table~\ref{tab:bestfit}. 
 Mass counts of passive galaxies in the different redshift bins are shown in Fig.~\ref{fig:counts}.

\subsection{The computation of the SMF} \label{sec:mf}

We computed the SMF through the non-parametric stepwise method of
\cite{castellano10b}, already applied in \cite{santini21} (see
references and details therein). We assume that the measured density
$\phi_i^{obs}$ in each mass bin $i$ can be described as the intrinsic
density $\phi_j^{true}$ in bin $j$ convolved with a transfer function
$S_{ij}$. 

The  transfer function 
is estimated from simulations  performed at the catalog level.  Mass counts in each subset (characterized by different photometric properties) are estimated by means of simulations performed ad-hoc and separately for the various fields or sub-fields. These simulations allow us to reconstruct the expected intrinsic (i.e. corrected) mass counts for a given area considering the noise properties of the relevant field. 

The simulations are based on \cite{bc03} synthetic templates and
were designed to cover the mass range between $10^{5.5}$ and
$10^{12.2} M_\odot$, to  mimic  
the average photometric noise and
the observed mass to light ratio.  More specifically, once perturbed with the noise properties of each field and fitted in the same way as real catalogs, mock galaxies were extracted to reproduce the observed mass to flux ratio in the $I$ band in different redshift and mass intervals. 
To avoid tuning our simulation to data sets that are appreciably affected by incompleteness, we only considered the deep fields, i.e. the Hubble Ultra Deep Field (HUDF)+HFF, for  the mass to flux ratio distribution 
at  $\log M/M_\odot$$<$8
  as well as at $\log  M/M_\odot$$=$8-9 and
  $z$$>$2.25, while mock galaxies with $\log M/M_\odot$$<$7 were tuned to the observed $\log M/M_\odot$$=$7-8  mass bin. 
  For passive and star-forming galaxies, the simulations are based on  templates satisfying the relevant {\it UVJ} criterion and are tuned to the observed mass to flux ratio distributions of the  corresponding population.

Simulated galaxies were  treated in
exactly the same way as the observed catalogues: they were selected if
 their SNR in the H160 band is larger than 5 
and were classified as passive and star-forming according to their
best-fit {\it UVJ}  colors.  For each field, the transfer function
$S_{ij}$ was constructed based on the number of simulated galaxies
that actually belong to bin $j$ but that, because of noise and
systematics, would be observed in bin $i$.  

The SMF was  finally
obtained
by inverting the linear system
$\phi_i^{obs}=\sum_j (S_{ij} \phi_j^{true})$. This technique takes
into account photometric scatter, mass uncertainties (i.e. percolation
of sources across adjacent bins), failures in the selection technique
and the Eddington bias, without any a-posteriori correction. 

 Redshift bins were chosen to overlap with the analysis of \cite{mcleod21}, for a cleaner comparison, and mass bins were chosen as a compromise between statistics and mass
resolution. 

Cosmic variance errors were added in quadrature following the prescriptions adopted in
\cite{santini21}.   Briefly, we used the QUICKCV code of \cite{moster11}, computing relative errors, for a given area, as a function of redshift and stellar mass. Following \cite{driver10}, we  reduced these relative errors by $\sqrt{N}$, where $N$ is the number of non-contiguous fields of similar area. We considered $N=5$ for the entire sample (due to the much larger area of each CANDELS field compared to the HFF) and $N=7$ when limiting the analysis to the deeper data (see below). 
Cosmic variance relative errors range from a few to 15\% at the highest redshift and masses. The resulting SMF over five redshift intervals are reported in Tables~\ref{tab:smf1} to \ref{tab:smf5}.

 Although the SMF is calculated by solving the linear system above and not correcting the observed densities for the fraction of galaxies that are missing, we can use our simulation to infer an estimate of the sample completeness. The latter is calculated as the fraction of simulated galaxies selected (after treating the simulation in the same way as the real data) in a given mass bin over the total number of input galaxies in the same  bin. The resulting completeness is shown in Fig.~\ref{fig:counts}. It is clear how the deepest fields show the higher completeness. In particular,  the HUDF and the HFF data  are $\sim$100\% complete out to M$\sim$$10^8M_\odot$ ($\sim$$10^9M_\odot$) at $z<1.5$ ($z>1.5$). 

We restricted our analysis to bins where the completeness
is above the 75\% level, in order to  contain potential systematics due
to the correction for incompleteness.  This effectively limits the bulk of the sample to higher significance, with the SNR distribution peaking at $\sim$10 (instead of 5). 
As a result of this further cut,  the SMF in most of the lowest
mass bins was estimated only from the deeper  and more complete HFF and HUDF data.  Figure~\ref{fig:counts} shows how the inclusion of the HFF data set is fundamental to probe the low mass regime of the SMF of passive galaxies. While relatively large numbers of candidates are observed in CANDELS, this sample is highly incomplete with respect to this galaxy population at stellar masses  below M$\sim$$10^{8.5}M_\odot$ ($\sim$$10^{9}M_\odot$) at $z<1$ ($z>1$). On the other side, the HUDF alone does not provide sufficient statistics to observe the rare low mass passive galaxies above $z\sim 1.5$.

The stepwise points were  fit with a single ($\phi
  = \ln(10) \times 10^{\phi^*_1} \times 10^{ (\mu-\mu^*) 
    (\alpha_1+1)} \times \exp(-10^{(\mu-\mu^*)})$)
and/or double ($ \phi
  = \ln(10) \times \left[10^{\phi^*_1} \times 10^{ (\mu-\mu^*) 
    (\alpha_1+1)} + 10^{\phi^*_2} \times 10^{ (\mu-\mu^*) 
    (\alpha_2+1)}\right] \times \exp(-10^{(\mu-\mu^*)})$)
Schechter function \citep{weigel16},  where $\mu=\log M$ and $\mu^*=\log M^*$.  For passive galaxies at $z\sim 2.5$, given the poor sampling in the low mass regime, we fix the faint-end slope to the best-fit value in the previous redshift bin. The best-fit parameters are presented in 
Table~\ref{tab:bestfit}.

\begin{table}
\centering
\begin{tabular} {ccc}
\hline \hline 
\noalign{\smallskip} 
Field & Area [arcmin$^2$] & F160W 5$\sigma$ limit\\
\noalign{\smallskip} 
\hline 
\noalign{\smallskip} 
GOODS-S HUDF & 5.1 & 29.04 \\
GOODS-S  deep & 58.9 & 27.32 \\
GOODS-S ERS & 51.5 & 27.05 \\
GOODS-S wide & 53.6 & 26.44 \\
GOODS-N deep & 93.0 & 27.14 \\
GOODS-N wide &80.0 & 26.33 \\
UDS  & 201.7 & 26.46 \\
EGS & 206.0 & 26.60 \\
COSMOS & 216.0 &  26.69 \\
A2744 parallel & 5.0 & 28.10 \\
M0416 parallel & 5.0 & 28.21 \\
M0717 parallel & 6.5 & 27.75 \\
M1149 parallel & 5.3 & 27.92 \\
AS1063 parallel & 6.6 & 27.63 \\
A370 parallel & 5.0 & 27.94 \\
\noalign{\smallskip} \hline \noalign{\smallskip}
\end{tabular}
\caption{Average 5$\sigma$ limiting total magnitudes 
  in the various fields (or sub-areas of the same field with inhomogeneous coverage), with associated areas.  The
limiting magnitude was calculated as the median magnitude of
sources with SNR between 4.8 and 5.2. }
\label{tab:depths}
\end{table}

\section{Results} \label{sec:results}

\begin{figure*}[ht!]
    \centering
\includegraphics[width=0.7\columnwidth,clip,viewport=320 25 555
750,angle=270]{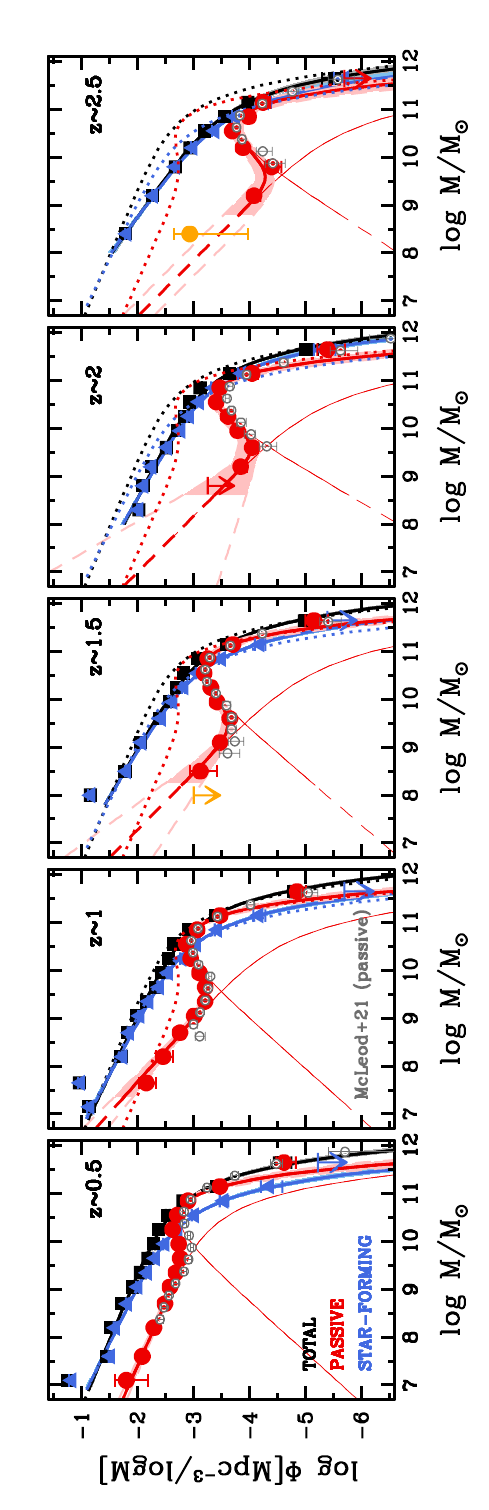}
\caption{Stellar Mass Function of all (black squares), passive (red solid circles) and
  star-forming (blue triangles) galaxies.  Upper limits are calculated at 1$\sigma$. Solid thick curves show the Schechter fits
  and the colored shaded areas represent the regions at 68\%
  confidence level considering the joint probability distribution
  function of the Schechter parameters. Dashed curves indicate the
  extrapolation of the Schechter fits to lower masses. 
  Thin curves are the two Schechter components  of the SMF of passive galaxies.   In each panel,
  dotted curves show the SMF at $z\sim0.5$.  Orange symbols are the results obtained at low masses for passive galaxies using HUDF only (see text), not included in the fits. Gray open circles are the
   SMF  of passive galaxies of \cite{mcleod21} in the same redshift bins.}
              \label{fig:mf}
\end{figure*}

\begin{figure}[!b]
    \centering
\includegraphics[width=9cm]{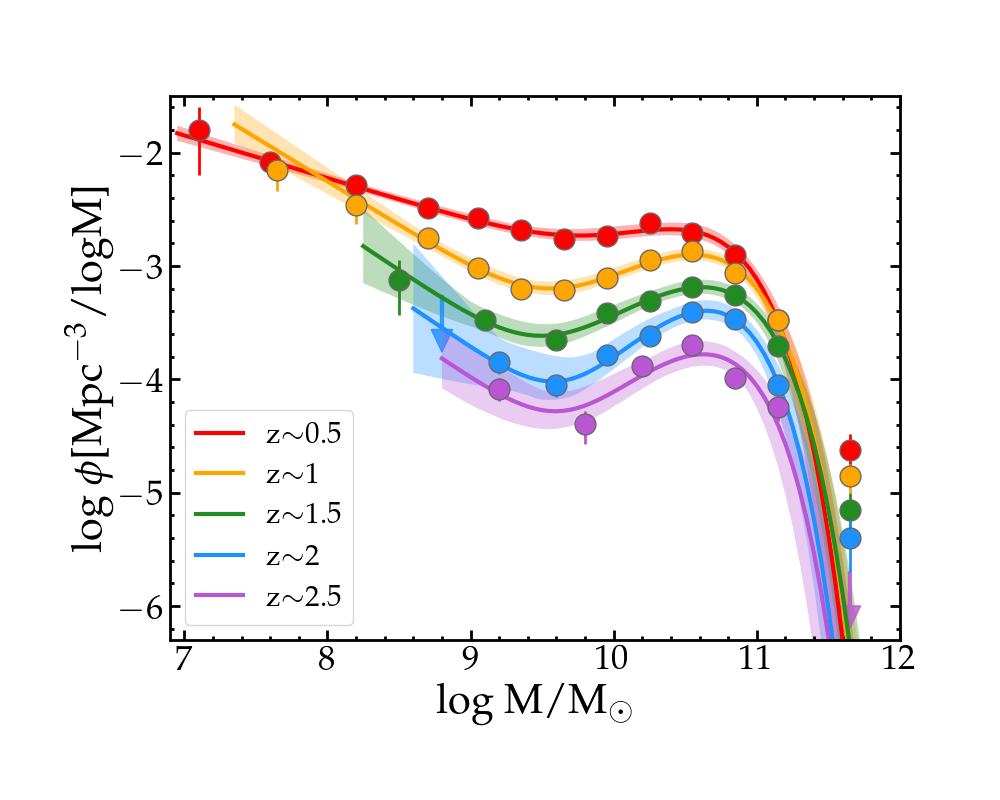}
  \caption{Evolution of the  SMF  of passive galaxies from $z\sim 2.5$ to $z\sim
    0.5$, color-coded as in the legend.  Points are the stepwise estimates, solid thick curves show the Schechter fits 
  and the colored shaded areas represent the regions at 68\%
  confidence level considering the joint probability distribution
  function of the Schechter parameters.
    }
              \label{fig:mfevol}
 \end{figure}

\subsection{The evolution of the  SMF  of passive galaxies }

We show in Fig.~\ref{fig:mf} the SMF of the total sample, of passive
and star-forming galaxies, as well as their fits with a Schechter function.  The  SMF of
passive and star-forming galaxies were fit with double and single
Schechter functions, respectively. As for the total  galaxy population, we compared
the single and double Schechter shapes, finding no significant
difference in the fits and in the final results. Thus we decided to
adopt the parameterization with less free parameters.  
 We note that the upper limit measured in the lowest mass bin of the SMF of passive galaxies at $z\sim2$ results from an almost complete lack of counts in  that redshift and mass range in the HUDF+HFF sample (see Fig.~\ref{fig:counts}). This can be ascribed to a combination of small statistics and cosmic variance. Although many more passive galaxies are selected in the same interval in the deep and wide CANDELS fields, we did not used  them due to the high incompleteness level of the parent sample. 

We also note that the exact low-mass slope of the SMF of passive galaxies may vary as a consequence of sample selection and mass binning choice, especially in the highest redshift bins, and will be better constrained with deeper JWST observations. However, the presence of a low mass population of passive galaxies is a solid result at all redshifts probed.  
As a matter of fact, the $10^9-10^{9.2}M_\odot$ points of the SMF are above the extrapolation of  a single Schechter fit to the high mass points by 0.5-0.8 dex and inconsistent with it at a 4-8$\sigma$ confidence level, with these deviations and inconsistencies increasing as the stellar mass decreases.

\begin{figure}[!t]
    \centering
  \includegraphics[width=1\columnwidth,clip,viewport=-20 -10 580 580]{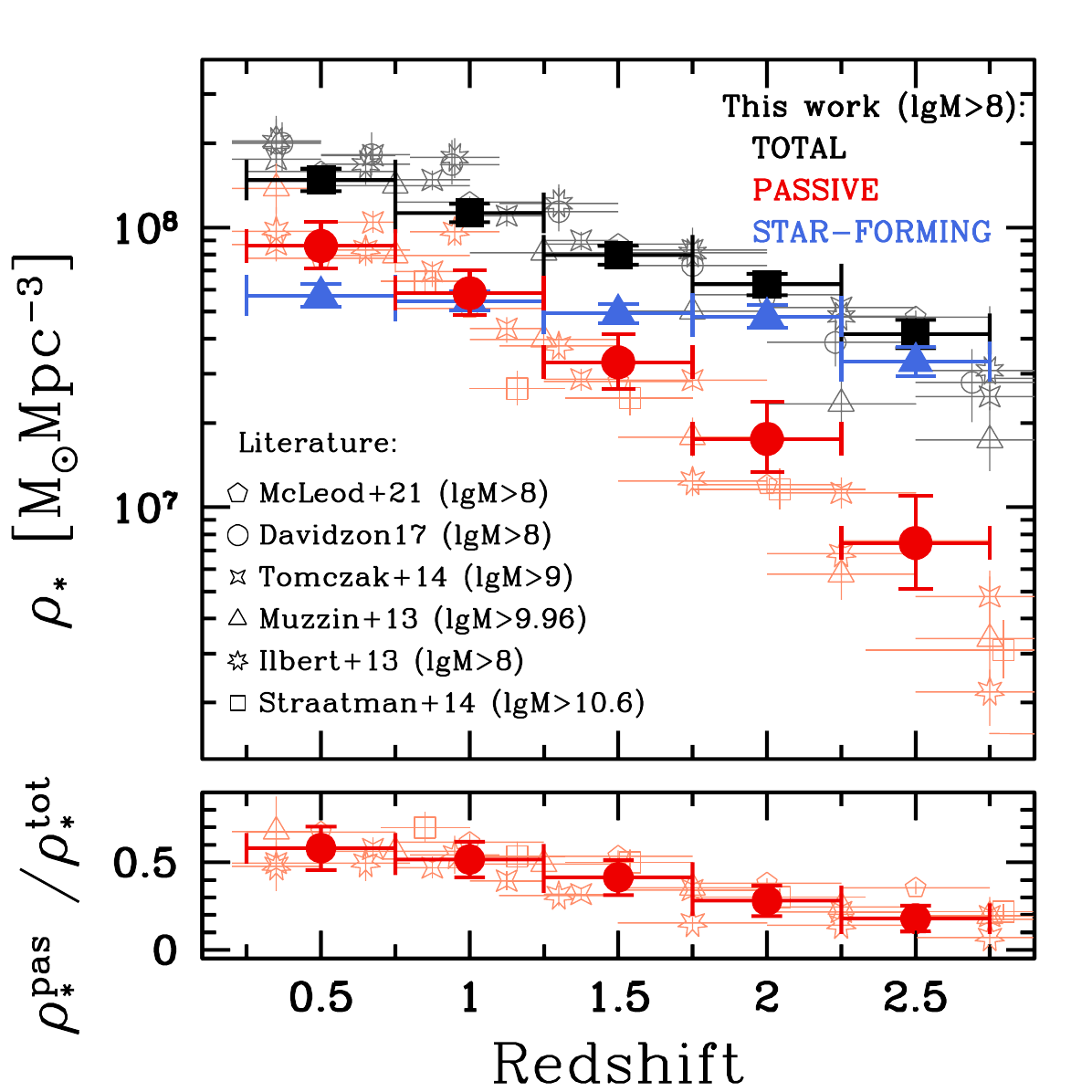}
  \caption{{\it Upper panel:} evolution of the Stellar Mass Density at
    $M>10^8M_\odot$ of all (solid black squares), passive (solid red
    circles) and star-forming (solid blue triangles) galaxies. Open
    black and reddish symbols show previous results from the literature,
    according to the legend, for all and passive galaxies,
    respectively. {\it Lower panel:} ratio of the SMD of passive
    galaxies to the overall galaxy population at $M>10^8M_\odot$,
    compared to the literature (symbols are as in the upper panel).  }
              \label{fig:md}
 \end{figure}

While the  SMF of star-forming and  total galaxies only evolve mildly over the
redshift range probed by the present work, the strong evolution
observed for the passive  sample suggests a  continuous and significant build-up for this
galaxy population over the last $\sim$11 Gyr \citep[see also][and
references therein]{santini21}.  Figure~\ref{fig:mfevol} shows the five
redshift bins on the same panel to better visualize the evolution in
the SMF  of passive galaxies, amounting to more than an order of magnitude from $z\sim2.5$
to $z\sim0.5$  at fixed stellar mass.  As might be expected, the Stellar Mass Density (SMD),
obtained by integrating the best-fit Schechter functions above
$10^8M_\odot$, shows a similar evolutionary trend
(Fig.~\ref{fig:md}). Passive galaxies account for 60\% of the total
mass in the nearby Universe, but for only $\sim$20\% at $z\sim2.5$, in
agreement with previous results
\citep{muzzin13,ilbert13,straatman14,tomczak14,mortlock15,davidzon17,mcleod21}. Over
the redshift interval probed by our analysis, the SMD of passive
galaxies increases by more than 
a factor of 10  (within the range 6-20 at 1$\sigma$), compared to a factor
of $\sim$3 and $\sim$1.5 evolution experienced by the total galaxy
population and the star-forming subsample, respectively.

 As discussed above, we observe clear evidence of an upturn at low stellar masses in the
SMF  of the passive galaxy sample at all redshifts probed by our analyses. 
This upturn has already been  identified in the local
Universe \citep[e.g.,][]{baldry12} and at $z\lesssim1.5$
\citep{tomczak14,mortlock15,davidzon17,mcleod21}, but was never observed before
at such high redshifts.  We compared our SMF with the recent
results of \cite{mcleod21} (Fig.~\ref{fig:mf}), which include CANDELS
data, along with ground-based surveys, and a correction for
incompleteness in the selection of passive galaxies, though with a
different technique. We found good agreement in the common mass range,
even at the high-mass end despite the much larger area probed by their
work. The inclusion of the deep HFF data, however, allow us to push
the analysis below $\log M/M_\odot$$\sim$9.5 at
$z>1.5$ and to probe
the existence of a secondary population of passive galaxies arising at
low masses already at these early epochs. 
 Indeed, as shown by Fig.~\ref{fig:counts} and discussed in Sect.~\ref{sec:mf}, the above mass range  is only poorly sampled by  the deepest CANDELS data,   with the HUDF area suffering from much lower statistics and larger cosmic variance than the full  HUDF+HFF deep sample used in this work.

\begin{figure}[!t]
    \centering
  \includegraphics[width=9cm]{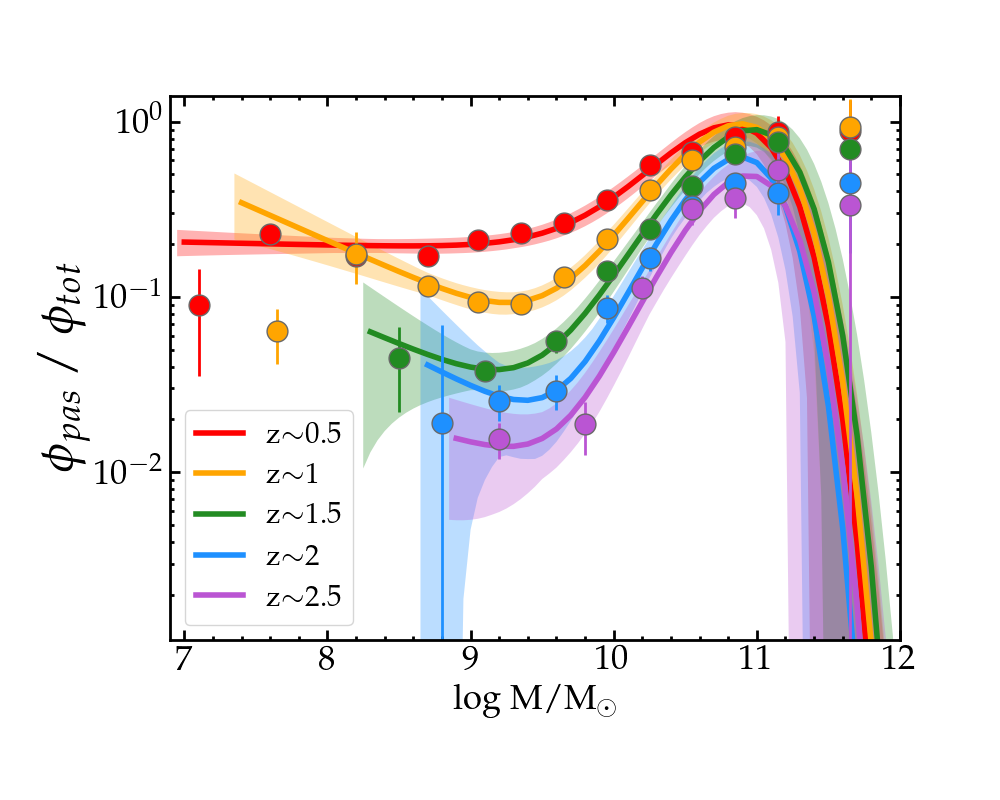}
  \caption{Ratio of the  SMF of passive to total galaxies at different
    redshifts, color-coded as in the legend. Symbols and curves show
    the ratio of the stepwise points and of the Schechter fits,
    respectively. }
              \label{fig:ratio}
 \end{figure}

 Despite the poor statistics, in the attempt of investigating even lower masses, we also calculated the SMF of passive galaxies from the HUDF alone  exploiting its higher depth  compared to the HFF program. As can be seen on Fig.~\ref{fig:counts}, the completeness in HUDF is indeed higher than for HFF. The difference is more pronounced at $z\sim2.5$, likely as a consequence of deeper IRAC photometry allowing a more accurate {\it UVJ} selection at high redshift. When the completeness is above the chosen threshold and the sparse HUDF counts allow us to estimate the SMF in a mass regime not probed by the combination of HUDF+HFF, we show the stepwise results on Fig.~\ref{fig:mf} as orange symbols. Although we do not use these points in the fits because of the poor statistics and large field-to-field variation, they corroborate the existence of a low mass population of passive galaxies at the highest redshifts probed by our analysis.

\begin{figure*}[t!]
    \centering
\includegraphics[width=0.7\columnwidth,clip,viewport=320 25 555
750,angle=270]{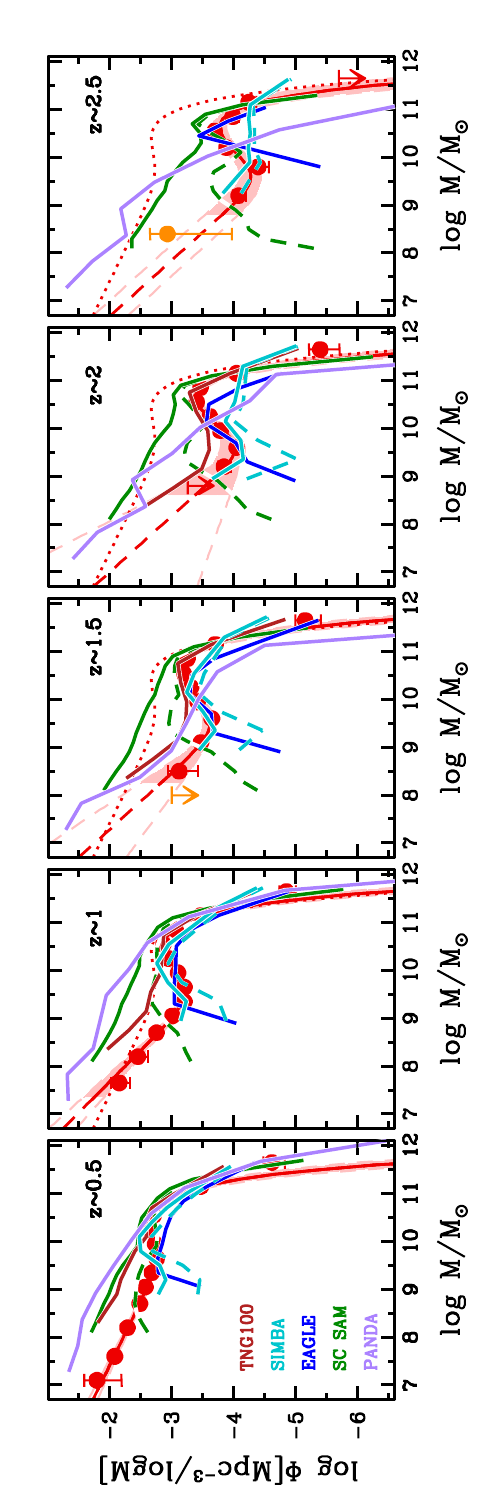}
\caption{Stellar Mass Function of passive galaxies (in red, as shown in
  Fig.~\ref{fig:mf}) compared to a set of theoretical predictions (see
  text and legend).  Dashed lines show the Santa Cruz (green) and SIMBA (light blue) models  after satellite galaxies have been removed.  Theoretical predictions are only shown above their mass completeness limit (see text), and predictions of zero
  galaxies have not been plotted. }
              \label{fig:mfmod}
\end{figure*}

 We run several simulations to rule out the possibility that the low-mass turnover is caused by a population of dusty star-forming galaxies misinterpreted as passive. 
Firstly, we produced a mock catalog of galaxies reproducing the photometric properties of one of the deep fields (we considered A2744), on which the estimate of the SMF at low masses is mostly based. We considered models with E(B-V)  in the range 0.5-1.1, normalized them to a H160 magnitude from 25 to 28, perturbed them with noise and fitted them with the same stellar library adopted to calculate the physical properties of our galaxies. We estimated the contamination as a function of stellar mass, and found  a roughly mass-independent value of 10-30\%  (we checked that lower values of extinction cause only a few per cent contamination  at $z\lesssim 1.5$). On the basis of these results, starting from the observed number of
passive  galaxies and dusty   ones (with different levels of extinction), 
we estimated an effective contamination ranging from a few to 10\% level at $M<10^9M_\odot$ and $\sim20-40\%$ at intermediate masses ($M\sim10^9-10^{10}M_\odot$), due to the lower number of passive galaxies in this mass range. 
Secondly, we considered the observed galaxy sample in A2744 independently on their dust obscuration, replicated it 10 times to increase the statistics, perturbed its photometry according to the typical noise properties of that field and fitted it in the usual way. 
We found a contamination of the order of 10\% at $z<2$ and 20-30\% at higher redshift. These levels of contamination, while potentially affecting the exact value of the slope of the SMF, do not change our main conclusion. Moreover, according to these tests, the contamination from dusty galaxies would make the slope of the low-mass component shallower rather than steeper.

\subsection{Two populations of passive galaxies}

The shape of the SMF and its evolution point to the existence of two
different populations of passive galaxies down to $z\sim 2.75$.  
The abundances of low- and high-mass passive galaxies evolve at markedly different rates: from $z\sim2.5$ to $z\sim0.5$ we observe an evolution by a factor of $\sim$20 and $\sim$6 at $\log M/M_\odot$$\sim$9 and $\sim 11$, respectively. 
A similar behaviour is shown by the Schechter fits: the normalization of the high- and low-mass components,  $\phi_{*1}$ and $\phi_{*2}$, evolve by one and two orders of magnitude, respectively (see Table~\ref{tab:bestfit}).

To assess the mutual importance of the two quenching modes  expected to be at work in the low and high mass regime, it is interesting to evaluate the fraction of passive galaxies
as a function of time and stellar mass. To this aim,
Fig.~\ref{fig:ratio} shows the ratio of the  SMF of passive to total galaxies
at each given redshift. Compared to the total population, passive
galaxies underwent a differential evolution with stellar mass: while
the majority of massive ($\log
M/M_\odot$$>$11-11.5) systems were already passive at
$z$$\sim$  2, and almost 50\% of them even at $z$$\sim$ 2.5, 
 the fraction of low mass ($\log
M/M_\odot$$<$10) passive galaxies experienced more than a factor of 10
evolution between $z\sim 2-2.5$ and the present epoch. 
 These findings, discussed in the next section,  qualitatively agree with previous studies at similar redshifts \citep[e.g.][]{fontana09,ilbert13,muzzin13,tomczak14,mortlock15,mcleod21}.

\section{Discussion} \label{sec:disc}

 In empirical and theoretical models, the evolution of the passive population is driven by a set of physical effects. In the  empirical model of \cite{peng10}, 
 the two populations arise as a result of two different
quenching processes: the mass-quenching mode, dominant at high masses,
and environmental-quenching, responsible for the upturn characterizing
the low mass end.

Massive galaxies reside in biased regions of the density field, have
started forming their stars earlier and have become passive at an
earlier time, in agreement with a downsizing scenario
\citep[e.g.][]{cowie96,fontanot09}. Moreover, cooling is inefficient
in high mass systems, hampering gas re-accretion and new star
formation episodes once the galaxy has run out of gas. Conversely, gas
is continuously expelled and re-accreted by low mass galaxies in the
field thanks to rapid and efficient cooling, giving rise to a more
prolonged star formation period and delayed quenching.
 However,  low-mass galaxies are quenched by  environmental effects \citep[e.g.][]{fossati17,papovich18,xie20}, as they enter in  the  virialized structures that progressively came into place
\citep[$z\sim2$, e.g.][]{overzier16}. This results in a differential
evolution of the SMF of passive galaxies
due to the  time-evolving role of the environment in suppressing the star
formation, which becomes increasingly more important at later
epochs. 
Similar conclusions on the evolution of environmental quenching processes are supported by both theoretical models \citep{vandevoort17} and observations \citep{kawinwanichakij17}.

This differential evolution is exactly the central finding of this work, as shown in  Figs.~\ref{fig:mfevol} and \ref{fig:ratio}. We cannot demonstrate that the low-mass part of the SMF is dominated by satellite galaxies in dense environments, as this would require an identification of the environment in which  each galaxy reside,  which is beyond the goal of this work. However, we  do find that the SMF of massive galaxies is largely in place already at $z\simeq 2$, while the low--mass part grows in subsequent times, lending support to the environmental explanation for the growth of the low-mass side.

\subsection{Comparison with theoretical predictions}

 To further explore the origin of the SMF of passive galaxies and the  connection of the low-mass upturn with the environment, we have performed a 
 comparison with theoretical predictions. In Fig.~\ref{fig:mfmod} we compare our results with several models:
the semianalytic model PANDA \citep{menci19}, the mock lightcones
extracted from the semianalytic Santa Cruz model \citep{somerville21},
and three hydrodynamic simulations, namely IllustrisTNG100
\citep{pillepich18a,nelson19}, EAGLE \citep{schaye15} and SIMBA
\citep{dave19}. 
Passive galaxies in the models were selected from the
simulated rest-frame $U$, $V$ and $J$ magnitudes, adopting the very same
cuts used on the data. TNG100 does not provide this information at
$z>2$, so we did not include it in our highest redshift bin.  
We plot the model predictions above their mass completeness limits. The PANDA model does not suffer from incompleteness, as the dark matter halo merging histories are based on a Monte Carlo procedure and go down to $10^5M_\odot$. The Santa Cruz model is complete  above $2.2 \times 10^{10}M_\odot$ in halo mass; following \cite{somerville21}, we only show the SMF above $10^8M_\odot$. The minimum stellar mass of a resolved galaxy in TNG100, EAGLE and SIMBA is $1.4 \times 10^8$, $2 \times 10^8$ and $5.8 \times 10^8 M_\odot$, respectively (for TNG100 we used the highest resolution simulation, i.e. TNG100-1).

In general, we see from Fig.~\ref{fig:mfmod} that models hardly reproduce the population of passive galaxies beyond the
local Universe, as also seen at even higher redshift in
\cite{santini21}.  The various predictions correctly model certain
redshift or mass intervals, including simulating the low and high mass
populations; in particular, SIMBA, and  to a lesser extent TNG100, well reproduce the
low-mass upturn at $z\sim 1.5-2$. However, none of the models are able
to match the observations over the entire redshift and stellar mass
dynamical range probed, suggesting that the physical processes
involved with the formation of passive galaxies are still not well
understood.

Hydrodynamical simulations may underpredict the faint end of the SMF
depending on the adopted mass resolution, yielding to zero galaxies
below a given threshold and truncated SMF (TNG100 makes an exception,
thanks to its slightly higher resolution and the adoption of the
moving-mesh method for solving the hydrodynamical equations,
\citealt{pillepich18a}). Additionally, an inefficient feedback
description is likely responsible for the lack of passive sources in
EAGLE, 
in particular 
at the lowest masses,  as well as in SIMBA at intermediate masses and $z\geq 1.5$.

Conversely, a common flaw of semianalytic models is their
overproduction of low mass passive galaxies as a consequence of the
introduction of efficient mechanisms to suppress excessive star
formation and mass production at high redshift.  
The overestimate of the number of passive
sources can be ascribed to the implementation  of gas
stripping in satellite galaxies, which is usually described as an
instantaneous process in the models (but see \citealt{henriques17} for
a partial solution to this problem), resulting in an unrealistically
rapid process.  This therefore invokes the need for improving the
description of the environmental processes in the models, in
particular in satellite galaxies (see also \citealt{calabro22}).

 From the Santa Cruz model and from the SIMBA simulation we also extracted information on the environment in which each galaxy resides. Following \cite{somerville21}, we identified satellites by selecting the galaxies flagged as belonging to a dark matter sub-halo, i.e. a halo that have become subsumed within another virialized halo and that is tidally stripped as it orbits within its host halo. As for SIMBA, we used the flag identifying the highest stellar mass galaxy as central, and all other galaxies in the same halo as satellites. 
With this information, we estimated the SMF of central galaxies only. 
It is clear from Fig.~\ref{fig:mfmod}  (dashed lines)  that the 
predicted SMF of passive galaxies at high masses drops when satellite galaxies are removed, consistently with the results of \cite{donnari21}  
(we note however that this effect is milder in SIMBA at $z\geq1.5$, likely due to resolution issues at the lowest masses and lower efficiency feedback processes in the satellites at  $z\sim 2.5$ relative to centrals). 
Since environmental effects such as ram-pressure or tidal stripping are expected to operate on low-mass satellites, on the basis of both observations \citep[e.g.][]{kovacs14,tal14,wetzel15,papovich18} and simulations \citep{fillingham18,samuel22}, this result supports the interpretation of the environment as major contributor in shaping the SMF of passive galaxies at low masses.

\section{Summary and conclusions}\label{sec:summary}

Reliable observations over a wide range of masses are crucial to
constrain the delicate processes responsible for the formation of
passive galaxies and improve their description in the models, in order to advance our understanding of the galaxy evolution
scenario.  Thanks to the inclusion of the deep HFF data in our sample,  as a complement to the CANDELS dataset, 
we were able to trace the evolution of the SMF at low masses up to
$z$$<$2.75.  

The observed shape of the SMF points to the existence of
two populations of passive galaxies that underwent different quenching
mechanisms and  are characterized by different evolutionary timescales.
Over the entire redshift range, and for the first time at $z$$>$1.5,
we observe the build-up of the low mass component, thought to
originate from environmental effects, efficient at suppressing the
star formation in low mass galaxies.

JWST observations  that are rapidly becoming available will allow the sampling of
the $\log M/M_\odot\lesssim 9$ regime at $z\gtrsim2$ with high
accuracy (less than a factor of 2 uncertainty on the stellar
mass). This will reduce the  current uncertainty on the  low-mass side of the SMF and
will allow us to push the analysis to even higher redshift. Accurate
observational results will allow us to better constrain the
theoretical description of the physical processes at play, with the
final aim of better understanding the role of environment, and its
evolution with cosmic time, in suppressing the star formation in
galaxies.

\begin{acknowledgments}
We thank the referee for the careful report that substantially improved the paper. We thank Rachel Somerville for useful discussions on the data/model
  comparison. The IllustrisTNG simulations were undertaken with
  compute time awarded by the Gauss Centre for Supercomputing (GCS)
  under GCS Large-Scale Projects GCS-ILLU and GCS-DWAR on the GCS
  share of the supercomputer Hazel Hen at the High Performance
  Computing Center Stuttgart (HLRS), as well as on the machines of the
  Max Planck Computing and Data Facility (MPCDF) in Garching,
  Germany. We acknowledge the Virgo Consortium for making their
  simulation data available. The EAGLE simulations were performed
  using the DiRAC-2 facility at Durham, managed by the ICC, and the
  PRACE facility Curie based in France at TGCC, CEA,
  Bruy\`eres-le-Ch\^atel. We also thank Romeel Dav\'e for help in
  using the SIMBA simulation data. We acknowledge  INAF Mini Grant 2022 "The evolution of passive galaxies through cosmic time".
\end{acknowledgments}

%

\vspace{5mm}






\begin{table}[h!]
\centering
\begin{tabular} {cccccccc} 
\hline \hline 
\noalign{\smallskip} 
\multicolumn{7}{c}{$0.25<z<0.75$}\\
\noalign{\smallskip} \hline \noalign{\smallskip}
$\log M$ & $N_{all}$ & $\log \phi_{all}$ & $N_{pas}$ & $\log\phi_{pas}$ & $N_{sf}$ & $\log\phi_{sf}$\\
\noalign{\smallskip} \hline \noalign{\smallskip}
7.10 & 294$^*$ & -0.75$_{-0.05}^{+0.04}$ &43$^*$ & -1.80$_{-0.40}^{+0.20}$ &251$^*$ & -0.82$_{-0.05}^{+0.04}$ \\
7.60 & 8730 & -1.44$_{-0.02}^{+0.02}$ &1165 & -2.08$_{-0.03}^{+0.03}$ &7565 & -1.51$_{-0.02}^{+0.02}$ \\
8.20 & 7952 & -1.52$_{-0.02}^{+0.02}$ &1363 & -2.29$_{-0.03}^{+0.02}$ &6589 & -1.60$_{-0.02}^{+0.02}$ \\
8.70 & 3360 & -1.72$_{-0.02}^{+0.02}$ &644 & -2.49$_{-0.03}^{+0.03}$ &2716 & -1.82$_{-0.02}^{+0.02}$ \\
9.05 & 1660 & -1.90$_{-0.03}^{+0.02}$ &343 & -2.58$_{-0.04}^{+0.03}$ &1317 & -2.00$_{-0.03}^{+0.02}$ \\
9.35 & 1195 & -2.05$_{-0.03}^{+0.02}$ &283 & -2.68$_{-0.04}^{+0.04}$ &912 & -2.17$_{-0.03}^{+0.03}$ \\
9.65 & 886 & -2.18$_{-0.03}^{+0.03}$ &233 & -2.76$_{-0.04}^{+0.04}$ &653 & -2.31$_{-0.03}^{+0.03}$ \\
9.95 & 718 & -2.28$_{-0.03}^{+0.03}$ &273 & -2.73$_{-0.04}^{+0.04}$ &445 & -2.49$_{-0.03}^{+0.03}$ \\
10.25 & 571 & -2.38$_{-0.03}^{+0.03}$ &308 & -2.62$_{-0.04}^{+0.03}$ &263 & -2.71$_{-0.04}^{+0.04}$ \\
10.55 & 398 & -2.54$_{-0.04}^{+0.03}$ &265 & -2.71$_{-0.04}^{+0.04}$ &133 & -3.01$_{-0.05}^{+0.04}$ \\
10.85 & 213 & -2.81$_{-0.04}^{+0.04}$ &173 & -2.90$_{-0.04}^{+0.04}$ &40 & -3.54$_{-0.08}^{+0.07}$ \\
11.15 & 53 & -3.42$_{-0.08}^{+0.07}$ &47 & -3.47$_{-0.08}^{+0.07}$ &6 & -4.36$_{-0.23}^{+0.15}$ \\
11.65 & 9 & -4.57$_{-0.18}^{+0.12}$ &8 & -4.62$_{-0.20}^{+0.14}$ &1 & $<$5.52 \\
\noalign{\smallskip} \hline \noalign{\smallskip}
\end{tabular}
\caption{Stellar Mass Function of all, passive and star-forming
  galaxies as calculated with the stepwise method in the
  $0.25<z<0.75$ redshift interval. The 1$\sigma$ uncertainties include Poissonian errors and cosmic variance. Stellar masses are in Solar masses and the SMF
  are in units of 1/Mpc$^3$/dex.  Columns 2, 4 and 6 indicate the number of sources in each bin. Asterisks denote the bins for which the SMF has been estimated from the HUDF+HFF data sets only to have a cleaner measurement thanks to the higher level of completeness.
}
\label{tab:smf1}
\end{table}

\begin{table}
\centering
\begin{tabular} {ccccccc} 
\hline \hline 
\noalign{\smallskip} 
\multicolumn{7}{c}{$0.75<z<1.25$}\\
\noalign{\smallskip} \hline \noalign{\smallskip}
$\log M$ & $N_{all}$ & $\log \phi_{all}$ & $N_{pas}$ & $\log\phi_{pas}$ & $N_{sf}$ & $\log\phi_{sf}$\\
\noalign{\smallskip} \hline \noalign{\smallskip}
7.15 & 311$^*$ & -1.13$_{-0.05}^{+0.05}$ &--&-- &308$^*$ & -1.14$_{-0.05}^{+0.05}$ \\
7.65 & 448$^*$ & -0.96$_{-0.04}^{+0.03}$ &18$^*$ & -2.16$_{-0.18}^{+0.13}$ &430$^*$ & -0.97$_{-0.04}^{+0.03}$ \\
8.20 & 10174 & -1.70$_{-0.02}^{+0.02}$ &21$^*$ & -2.46$_{-0.17}^{+0.12}$ &9623 & -1.74$_{-0.02}^{+0.02}$ \\
8.70 & 5912 & -1.82$_{-0.02}^{+0.02}$ &448 & -2.76$_{-0.03}^{+0.03}$ &5464 & -1.85$_{-0.02}^{+0.02}$ \\
9.05 & 2978 & -1.98$_{-0.02}^{+0.02}$ &277 & -3.01$_{-0.04}^{+0.04}$ &2701 & -2.03$_{-0.02}^{+0.02}$ \\
9.35 & 2014 & -2.16$_{-0.02}^{+0.02}$ &189 & -3.21$_{-0.05}^{+0.05}$ &1825 & -2.20$_{-0.02}^{+0.02}$ \\
9.65 & 1425 & -2.32$_{-0.02}^{+0.02}$ &205 & -3.21$_{-0.05}^{+0.04}$ &1220 & -2.38$_{-0.02}^{+0.02}$ \\
9.95 & 1125 & -2.44$_{-0.02}^{+0.02}$ &266 & -3.10$_{-0.04}^{+0.04}$ &859 & -2.54$_{-0.03}^{+0.02}$ \\
10.25 & 833 & -2.56$_{-0.03}^{+0.03}$ &358 & -2.95$_{-0.04}^{+0.03}$ &475 & -2.79$_{-0.03}^{+0.03}$ \\
10.55 & 660 & -2.65$_{-0.03}^{+0.03}$ &394 & -2.87$_{-0.03}^{+0.03}$ &266 & -3.04$_{-0.04}^{+0.03}$ \\
10.85 & 363 & -2.92$_{-0.03}^{+0.03}$ &250 & -3.06$_{-0.04}^{+0.04}$ &113 & -3.42$_{-0.05}^{+0.04}$ \\
11.15 & 118 & -3.39$_{-0.06}^{+0.05}$ &98 & -3.48$_{-0.06}^{+0.05}$ &20 & -4.16$_{-0.12}^{+0.10}$ \\
11.65 & 11 & -4.82$_{-0.18}^{+0.12}$ &10 & -4.85$_{-0.15}^{+0.11}$ &1 & $<$6.00 \\
\noalign{\smallskip} \hline \noalign{\smallskip}
\end{tabular}
\caption{Same as Table~\ref{tab:smf1} for the 
  $0.75<z<1.25$ redshift interval. 
}
\label{tab:smf2}
\end{table}

\begin{table}
\centering
\begin{tabular} {ccccccc} 
\hline \hline 
\noalign{\smallskip} 
\multicolumn{7}{c}{$1.25<z<1.75$}\\
\noalign{\smallskip} \hline \noalign{\smallskip}
$\log M$ & $N_{all}$ & $\log \phi_{all}$ & $N_{pas}$ & $\log\phi_{pas}$ & $N_{sf}$ & $\log\phi_{sf}$\\
\noalign{\smallskip} \hline \noalign{\smallskip}
8.00 & 322$^*$ & -1.16$_{-0.05}^{+0.04}$ &--&-- &319$^*$ & -1.16$_{-0.05}^{+0.04}$ \\
8.50 & 10613 & -1.77$_{-0.02}^{+0.02}$ &6$^*$ & -3.12$_{-0.31}^{+0.18}$ &10421 & -1.80$_{-0.02}^{+0.02}$ \\
9.10 & 6781 & -2.05$_{-0.02}^{+0.02}$ &232 & -3.47$_{-0.04}^{+0.04}$ &6549 & -2.07$_{-0.02}^{+0.02}$ \\
9.60 & 2106 & -2.40$_{-0.02}^{+0.02}$ &135 & -3.65$_{-0.06}^{+0.06}$ &1971 & -2.42$_{-0.02}^{+0.02}$ \\
9.95 & 1095 & -2.57$_{-0.03}^{+0.02}$ &167 & -3.42$_{-0.05}^{+0.05}$ &928 & -2.63$_{-0.03}^{+0.02}$ \\
10.25 & 791 & -2.70$_{-0.03}^{+0.03}$ &211 & -3.31$_{-0.05}^{+0.04}$ &580 & -2.82$_{-0.03}^{+0.03}$ \\
10.55 & 585 & -2.82$_{-0.03}^{+0.03}$ &268 & -3.19$_{-0.04}^{+0.04}$ &317 & -3.08$_{-0.04}^{+0.03}$ \\
10.85 & 326 & -3.07$_{-0.04}^{+0.03}$ &203 & -3.25$_{-0.04}^{+0.04}$ &123 & -3.50$_{-0.05}^{+0.05}$ \\
11.15 & 97 & -3.59$_{-0.06}^{+0.06}$ &72 & -3.71$_{-0.07}^{+0.06}$ &25 & -4.18$_{-0.10}^{+0.08}$ \\
11.65 & 9 & -5.00$_{-0.15}^{+0.11}$ &7 & -5.15$_{-0.24}^{+0.15}$ &2 & $<$5.70 \\
\noalign{\smallskip} \hline \noalign{\smallskip}
\end{tabular}
\caption{Same as Table~\ref{tab:smf1} for the 
  $1.25<z<1.75$ redshift interval. 
}
\label{tab:smf3}
\end{table}

\begin{table}
\centering
\begin{tabular} {ccccccc} 
\hline \hline 
\noalign{\smallskip} 
\multicolumn{7}{c}{$1.75<z<2.25$}\\
\noalign{\smallskip} \hline \noalign{\smallskip}
$\log M$ & $N_{all}$ & $\log \phi_{all}$ & $N_{pas}$ & $\log\phi_{pas}$ & $N_{sf}$ & $\log\phi_{sf}$\\
\noalign{\smallskip} \hline \noalign{\smallskip}
8.30 & 5822 & -2.02$_{-0.02}^{+0.02}$ &--&-- &5783 & -2.02$_{-0.02}^{+0.02}$ \\
8.80 & 4482 & -2.10$_{-0.02}^{+0.02}$ &1$^*$ & $<$-3.26 &4434 & -2.11$_{-0.02}^{+0.02}$ \\
9.20 & 3129 & -2.25$_{-0.02}^{+0.02}$ &44 & -3.84$_{-0.11}^{+0.09}$ &3085 & -2.25$_{-0.02}^{+0.02}$ \\
9.60 & 1790 & -2.52$_{-0.03}^{+0.02}$ &56 & -4.05$_{-0.10}^{+0.08}$ &1734 & -2.53$_{-0.03}^{+0.02}$ \\
9.95 & 826 & -2.72$_{-0.03}^{+0.03}$ &89 & -3.79$_{-0.08}^{+0.07}$ &737 & -2.77$_{-0.03}^{+0.03}$ \\
10.25 & 627 & -2.84$_{-0.03}^{+0.03}$ &115 & -3.62$_{-0.07}^{+0.06}$ &512 & -2.92$_{-0.03}^{+0.03}$ \\
10.55 & 517 & -2.92$_{-0.04}^{+0.04}$ &180 & -3.40$_{-0.05}^{+0.05}$ &337 & -3.10$_{-0.04}^{+0.04}$ \\
10.85 & 322 & -3.12$_{-0.04}^{+0.04}$ &136 & -3.47$_{-0.05}^{+0.05}$ &186 & -3.36$_{-0.05}^{+0.04}$ \\
11.15 & 91 & -3.65$_{-0.07}^{+0.06}$ &36 & -4.05$_{-0.10}^{+0.08}$ &55 & -3.86$_{-0.08}^{+0.07}$ \\
11.65 & 9 & -5.05$_{-0.18}^{+0.12}$ &4 & -5.40$_{-0.30}^{+0.18}$ &5 & -5.30$_{-0.22}^{+0.15}$ \\
\noalign{\smallskip} \hline \noalign{\smallskip}
\end{tabular}
\caption{Same as Table~\ref{tab:smf1} for the 
  $1.75<z<2.25$ redshift interval. 
}
\label{tab:smf4}
\end{table}

\begin{table}
\centering
\begin{tabular} {ccccccc} 
\hline \hline 
\noalign{\smallskip} 
\multicolumn{7}{c}{$2.25<z<2.75$}\\
\noalign{\smallskip} \hline \noalign{\smallskip}
$\log M$ & $N_{all}$ & $\log \phi_{all}$ & $N_{pas}$ & $\log\phi_{pas}$ & $N_{sf}$ & $\log\phi_{sf}$\\
\noalign{\smallskip} \hline \noalign{\smallskip}
8.40 & 547$^*$ & -1.78$_{-0.04}^{+0.04}$ &--&-- &540$^*$ & -1.81$_{-0.04}^{+0.04}$ \\
9.20 & 6002 & -2.27$_{-0.02}^{+0.02}$ &145 & -4.08$_{-0.11}^{+0.09}$ &5857 & -2.28$_{-0.02}^{+0.02}$ \\
9.80 & 1320 & -2.67$_{-0.03}^{+0.03}$ &34 & -4.40$_{-0.17}^{+0.12}$ &1286 & -2.68$_{-0.03}^{+0.03}$ \\
10.20 & 671 & -2.94$_{-0.03}^{+0.03}$ &90 & -3.89$_{-0.07}^{+0.06}$ &581 & -2.99$_{-0.04}^{+0.03}$ \\
10.55 & 280 & -3.20$_{-0.05}^{+0.05}$ &88 & -3.69$_{-0.07}^{+0.06}$ &192 & -3.37$_{-0.06}^{+0.05}$ \\
10.85 & 126 & -3.55$_{-0.06}^{+0.05}$ &46 & -3.99$_{-0.09}^{+0.08}$ &80 & -3.75$_{-0.07}^{+0.06}$ \\
11.15 & 45 & -3.97$_{-0.10}^{+0.08}$ &22 & -4.24$_{-0.12}^{+0.10}$ &23 & -4.26$_{-0.13}^{+0.10}$ \\
11.65 & 3 & -5.52$_{-0.48}^{+0.22}$ &1 & $<$6.00 &2 & -5.70$_{-0.30}^{+0.18}$ \\
\noalign{\smallskip} \hline \noalign{\smallskip}
\end{tabular}
\caption{Same as Table~\ref{tab:smf1} for the 
  $2.25<z<2.75$ redshift interval. 
}
\label{tab:smf5}
\end{table}

\begin{table*}
\centering
\begin{tabular} {cccccccc} 
\hline \hline 
\noalign{\smallskip} 
Redshift & N & $\log M^*/M_\odot$ & $\alpha_1$ & $\log \phi_{*1}$/Mpc$^{-3}$
  & $\alpha_2$ & $\log \phi_{*2}$/Mpc$^{-3}$ & $\log \rho_{>8}/(M_\odot$Mpc$^{-3}$)\\
\noalign{\smallskip} \hline \noalign{\smallskip}
\multicolumn{8}{c}{{\it All galaxies}}\\
\noalign{\smallskip} \hline \noalign{\smallskip}
0.25 - 0.75 & 26039 & 10.98 $\pm$ 0.03 & -1.36 $\pm$ 0.01 & -2.96 $\pm$ 0.03 &  -- &  -- & 8.17$_{-0.04}^{+0.04}$ \\
0.75 - 1.25 & 26372 & 11.08 $\pm$ 0.03 & -1.41 $\pm$ 0.01 & -3.20 $\pm$ 0.03 &  -- &  -- & 8.05$_{-0.03}^{+0.03}$ \\
1.25 - 1.75 & 22725 & 11.11 $\pm$ 0.03 & -1.50 $\pm$ 0.01 & -3.45 $\pm$ 0.04 &  -- &  -- & 7.90$_{-0.03}^{+0.03}$ \\
1.75 - 2.25 & 17615 & 11.10 $\pm$ 0.03 & -1.45 $\pm$ 0.01 & -3.50 $\pm$ 0.04 &  -- &  -- & 7.80$_{-0.04}^{+0.04}$ \\
2.25 - 2.75 & 8994 & 11.05 $\pm$ 0.06 & -1.61 $\pm$ 0.03 & -3.76 $\pm$ 0.07 &  -- &  -- & 7.62$_{-0.05}^{+0.05}$ \\
\noalign{\smallskip} \hline \noalign{\smallskip}
\multicolumn{8}{c}{{\it Passive galaxies}}\\
\noalign{\smallskip} \hline \noalign{\smallskip}
0.25 - 0.75 & 5148 & 10.54 $\pm$ 0.06 & -0.06 $\pm$ 0.18 & -2.67 $\pm$ 0.04 & -1.38 $\pm$ 0.03 & -3.54 $\pm$ 0.09 & 7.93$_{-0.08}^{+0.08}$ \\
0.75 - 1.25 & 2534 & 10.59 $\pm$ 0.05 & -0.04 $\pm$ 0.13 & -2.84 $\pm$ 0.03 & -1.77 $\pm$ 0.08 & -4.61 $\pm$ 0.18 & 7.77$_{-0.08}^{+0.08}$ \\
1.25 - 1.75 & 1301 & 10.62 $\pm$ 0.06 & -0.00 $\pm$ 0.18 & -3.12 $\pm$ 0.03 & -1.85 $\pm$ 0.31 & -5.21 $\pm$ 0.55 & 7.52$_{-0.10}^{+0.10}$ \\
1.75 - 2.25 & 661 & 10.48 $\pm$ 0.07 & 0.47 $\pm$ 0.30 & -3.37 $\pm$ 0.04 & -1.85 $\pm$ 0.64 & -5.34 $\pm$ 0.86 & 7.24$_{-0.12}^{+0.13}$ \\
2.25 - 2.75 & 426 & 10.52 $\pm$ 0.10 & 0.31 $\pm$ 0.29 & -3.73 $\pm$ 0.04 & -1.85 $\pm$ 0.00 & -5.64 $\pm$ 0.18 & 6.87$_{-0.16}^{+0.17}$ \\
\noalign{\smallskip} \hline \noalign{\smallskip}
\multicolumn{8}{c}{{\it Star-forming galaxies}}\\
\noalign{\smallskip} \hline \noalign{\smallskip}
0.25 - 0.75 & 20891 & 10.60 $\pm$ 0.04 & -1.42 $\pm$ 0.01 & -3.01 $\pm$ 0.04 &  -- &  -- & 7.76$_{-0.04}^{+0.04}$ \\
0.75 - 1.25 & 23305 & 10.72 $\pm$ 0.03 & -1.45 $\pm$ 0.01 & -3.18 $\pm$ 0.04 &  -- &  -- & 7.74$_{-0.03}^{+0.03}$ \\
1.25 - 1.75 & 21235 & 10.81 $\pm$ 0.04 & -1.53 $\pm$ 0.02 & -3.36 $\pm$ 0.05 &  -- &  -- & 7.69$_{-0.03}^{+0.03}$ \\
1.75 - 2.25 & 16868 & 11.04 $\pm$ 0.04 & -1.49 $\pm$ 0.02 & -3.59 $\pm$ 0.04 &  -- &  -- & 7.68$_{-0.04}^{+0.04}$ \\
2.25 - 2.75 & 8561 & 10.89 $\pm$ 0.07 & -1.63 $\pm$ 0.03 & -3.71 $\pm$ 0.08 &  -- &  -- & 7.52$_{-0.05}^{+0.05}$ \\
\noalign{\smallskip} \hline \noalign{\smallskip}
\end{tabular}
\caption{Best-fit parameters and their 1$\sigma$ uncertainties in the different redshift
  intervals derived from fitting the stepwise SMF with a single 
    or  double 
Schechter function for all galaxies and for the passive and
  star-forming populations.  The second column indicates the numbers of galaxies used for computing the SMF in each redshift bin. 
  The last column reports the corresponding mass density $\rho$
  obtained by integrating the SMF above $10^{8} M_\odot$. We note that
  for passive galaxies at $z\sim 2.5$,   $\alpha_2$ was fixed to the best-fit value in the previous redshift interval. 
}
\label{tab:bestfit}
\end{table*}



\bibliographystyle{aasjournal}



\end{document}